\documentclass[pra,twocolumn,showpacs,preprintnumbers,amsmath,amssymb]{revtex4-1}
\makeatletter
\def\@dotsep{4.5}
\makeatother
\usepackage[T1]{fontenc}
\usepackage[utf8]{inputenc}
\usepackage{amsmath,amssymb}
\usepackage{color}
\usepackage{graphicx}
\usepackage{amssymb}
\usepackage{color}
\usepackage{cancel}
\begin{document}


\title{Charged  Topological Solitons in Zigzag Graphene Nanoribbons}

\author{M. P. L\'opez-Sancho }
\affiliation{Instituto de Ciencia de Materiales de Madrid, CSIC, 28049 Cantoblanco, Spain}
\author{Luis Brey}
\affiliation{Instituto de Ciencia de Materiales de Madrid, CSIC, 28049 Cantoblanco, Spain}
\email{brey@icmm.csic.es}
\date{\today}
\pacs{}

\begin{abstract}
Graphene nanoribbons with zigzag terminated edges have a magnetic ground state
characterized by edge ferromagnetism  and antiferromagnetic inter edge coupling. 
This broken symmetry state is  degenerate in the spin orientation and we show that, 
associated with this degeneracy,  the system has topological solitons. 
The solitons 
appear at the interface between  degenerate ground states.
These solitons are the relevant charge excitations in the system.
When charge is added to the nanoribbon, the system energetically prefers to create magnetic domains and accommodate 
the extra electrons in the interface solitons rather than setting them in the conduction band.
\end{abstract}
\maketitle

\date{\today}%

{\it Introduction.}-  Graphene nanoribbons are very interesting systems not only for their potential applications  as
connectors in graphene based nano devices, but also by their fundamental physical properties.
Many of the exotic  properties of graphene have origin on the bipartite character of the honeycomb
lattice\cite{Guinea_2009,Katsnelson-book}. Similarly, the magnetic and electric properties of graphene nanoribbons depend dramatically on the atomic termination of the edges\cite{key-loc3,Wakabayashi:1999aa,Brey:2006aa,Palacios_2010,Zarea:2008aa}.  
The chiral nature of the low energy carriers in graphene makes zigzag nanoribbons to have highly 
degenerate zero energy states localized at the edges,
and this unique property has stimulated the study of magnetic instabilities in these ribbons\cite{Son:2006aa,Vozmediano:2005aa,Pisani:2007aa,Fernandez-Rossier:2007aa,Yazyev:2008aa,Yazyev:2010aa,Rhim:2009aa,Santos:2009aa,Guclu:2013aa}.

Zigzag graphene  nanoribbons (ZZGN's) are characterized by the number of atoms in the unit cell, $N_x$, that corresponds to  a ribbon width 
$W$=$\sqrt{3} N_x a/4$, being  $a$ the graphene lattice parameter. 
The momentum
of the electrons along the ribbon, $k$, is restricted to take values in the one-dimensional Brillouin zone, $0<k<\frac {2 \pi}a$. 
The number of atoms per unit cell determines the number of electronic bands per spin in 
the  Brillouin zone. 
The low energy conduction band and the high energy valence band are
degenerate at the center of the Brillouin zone, and these two bands becomes flatten as the width of the ribbon increases, so that 
they are practically degenerate at zero energy  in the range $\frac {2 \pi} {3a} < k< \frac {4\pi}{3a}$. These  zero energy states correspond to states localized at the 
edges of the nanoribbons and  because they are located on opposite sublattices of the graphene unit cell they are not coupled by the kinetic energy part of the Hamiltonian.

\begin{figure}[htbp]
\includegraphics[width=7.cm,clip]{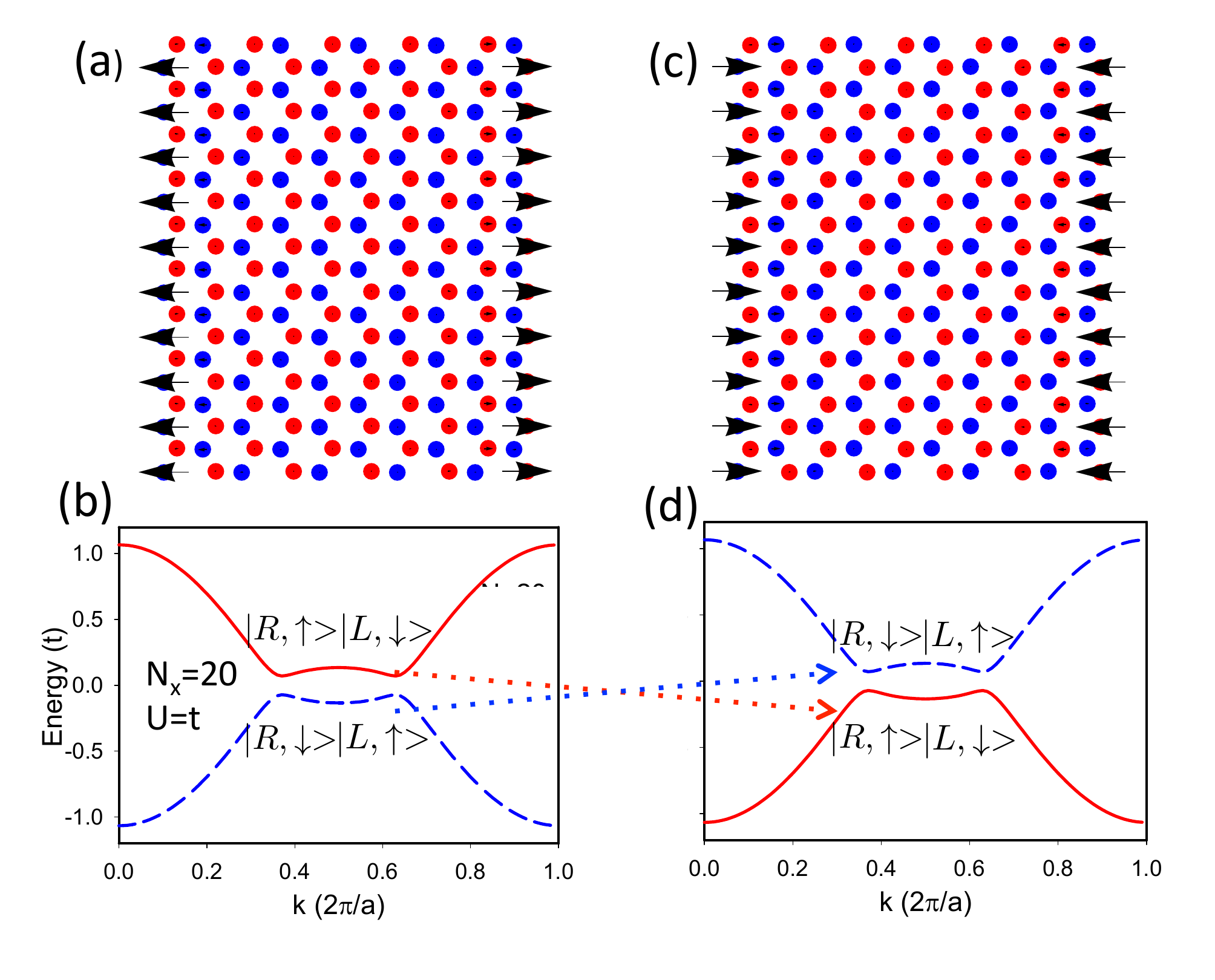}
\caption{(a) Schematic picture of the spin polarization of an undopped  zigzag nanoribbon. (b) Lowest energy conduction band and highest valence band of a ZZGN with $N$=20, and $U$=$t$. (c) spin polarization and (d) band structure for the state degenerate 
with the ground state shown in (a) and (b).
In (a) and (c) the spin orientations  are rotated  to make the figure clearer. } 
\label{Figure1}
\end{figure}

The degeneracy of the valence and conduction bands
produces a sharp peak in the density of states at the Fermi energy that
makes  the system unstable against broken symmetry ground states. 
Several {\it ab initio} density functional based calculations\cite{Son:2006aa, Pisani:2007aa,Yazyev:2008aa}, and tight-binding Hamiltonians with  long-range\cite{Jung:2011aa,2017arXiv170200452S} or
on-site\cite{Wakabayashi:1999aa} interactions have shown that the 
electron-electron interaction  opens a gap in the electronic structure and induces  magnetic order in the ground state. 
All the theoretical calculations indicate the existence of  spin polarization localized near each 
edge and an  antiferromagnetic coupling between opposite edges, see Fig.\ref{Figure1}(a).  
This antiferromagnetic coupling between opposite edges satisfies the Lieb's theorem\cite{Lieb:1989aa,Brey:2007aa}.
The exchange interaction between electrons favors the occupancy of electronic states in an edge with a spin orientation
and states with opposite spin in the opposite edge. 
Zigzag graphene nanoribbons have been obtained  by using top-down 
approximations\cite{Han:2007aa,Kosynkin:2009aa,Li:2008aa,Wang:2010aa,Magda:2014aa,Ma:2013aa}, 
by growing graphene epitaxially on silicon carbide\cite{SprinkleM.:2010aa,
Hicks:2013aa,Baringhaus:2014aa,Baringhaus:2016aa}  and by using on-surface syntesis techniques\cite{Ruffieux:2016aa}.
Some of these nanoribbons  show ballistic transport and have a high performance\cite{Baringhaus:2014aa,Baringhaus:2016aa}.
In addition, experiments have found proof of  magnetic order  at zigzag graphene nanoribbons.\cite{Tao:2011aa,Magda:2014aa,Ruffieux:2016aa}

\par
\noindent
{\it Main conclusions.}- By inverting the spin polarization of the full  system  there is  another energy degenerate ground state, 
see Fig.\ref{Figure1}(a)-(c).
The origin of the degeneracy is  the broken symmetry  in the spin sector that occurs in the ground state. 
The band structures of the degenerate ground states are inverted, see Fig.\ref{Figure1}(b)-(d),
and we argue that, when
connecting two domains with opposite mass, a symmetry-protected topological state will appear at the interface.
Here, the topological defects are soliton-like excitons that carry a charge $\pm e$, with half electron localized at each
edge of the  nanoribbon.  We claim that when doping, the extra charge will accommodate creating domain walls between opposite polarized degenerate ground states.
Interestingly the topological properties of zigzag graphene nanoribbons are generated by the electron electron interaction
and not by spin-orbit\cite{Kane-2005}, orbital\cite{Brey:2005aa} or bond ordering\cite{Heeger:1988aa}.
In the following of this letter, we develop these arguments and present numerical results supporting the 
existence of topological charged excitations in zigzag graphene nanoribbons.

\par
\noindent
{\it Hamiltonian}-
In this work we describe the electron-electron interaction in the on-site Hubbard model,
\begin{equation}
H=-t \sum _{<i,j>,\sigma} c^+_{i,\sigma} c _{j,\sigma}
+U \sum_{i} \hat n_{i,\uparrow} \hat n_{i,\downarrow} \, ,
\end{equation}
here $c^+_{i,\sigma}$ creates an electron at  site $i$ with spin $\sigma$
and $\hat n_{i,\sigma} = c ^+ _{i,\sigma} c_{i,\sigma}$. In this Hamiltonian hopping
exits between nearest neighbor $\pi$ orbitals with a value $t \approx 2.7$eV\cite{Guinea_2009,Katsnelson-book}.
The Hubbard  model takes into account the short-range part of the Coulomb interaction
through the parameter $U>0$.  Experiments\cite{Kuroda:1987aa, Thomann:1985aa}  give the range of values
$U \sim 3.0-3.5eV$. In this work we use a value of  $U \sim t$ that yields results in agreement with 
density functional theory\cite{Son:2006aa,Fernandez-Rossier:2007aa}.
Exchange interaction is the main ingredient for obtaining  magnetic order and therefore Hartree-Fock pairing of the  operators is a good
approximation for describing magnetic properties of graphene zigzag nanoribbons.
The unrestricted Hartree-Fock approximation for the  Hubbard term read 
\begin{widetext}
\begin{equation}
V_{mf} \!=\! U \sum _{i} \left ( \sum_{\sigma} \left( \hat  n_{i,\sigma} < \! \hat n_{i,-\sigma } \!> \!-c^+_{i,-\sigma} c_{i,\sigma}<\!c^+_{i,\sigma} c_{i,-\sigma}\!> \right )
-\!< \! n_{i,\uparrow }\!> <\! n_{i,\downarrow} \!>\!+\!<\!c^+_{i,\uparrow} c_{i,\downarrow}\!><\!c^+_{i,\downarrow} c_{i,\uparrow}\!>    \right )
\end{equation}
\end{widetext}
where
$<\hat O >$ means expectation value of the $\hat O$ operator.
By solving self-consistently the Hamiltonian we obtain the expectation value of charge and spin at every site of the nanoribbon
and  the band structure.
In  Fig.\ref{Figure1} we plot a typical band structure and magnetic order for the case of an undopped nanoribbon.
The bands are spin degenerate and electron electron interaction creates a magnetic order, with ferromagnetic order at the edges that are
anti-ferromagnetically  coupled.   Because graphene has a bipartite lattice atoms in  different sublattices have  opposite spin polarization.


\begin{figure}[htbp]
\includegraphics[width=7.cm,clip]{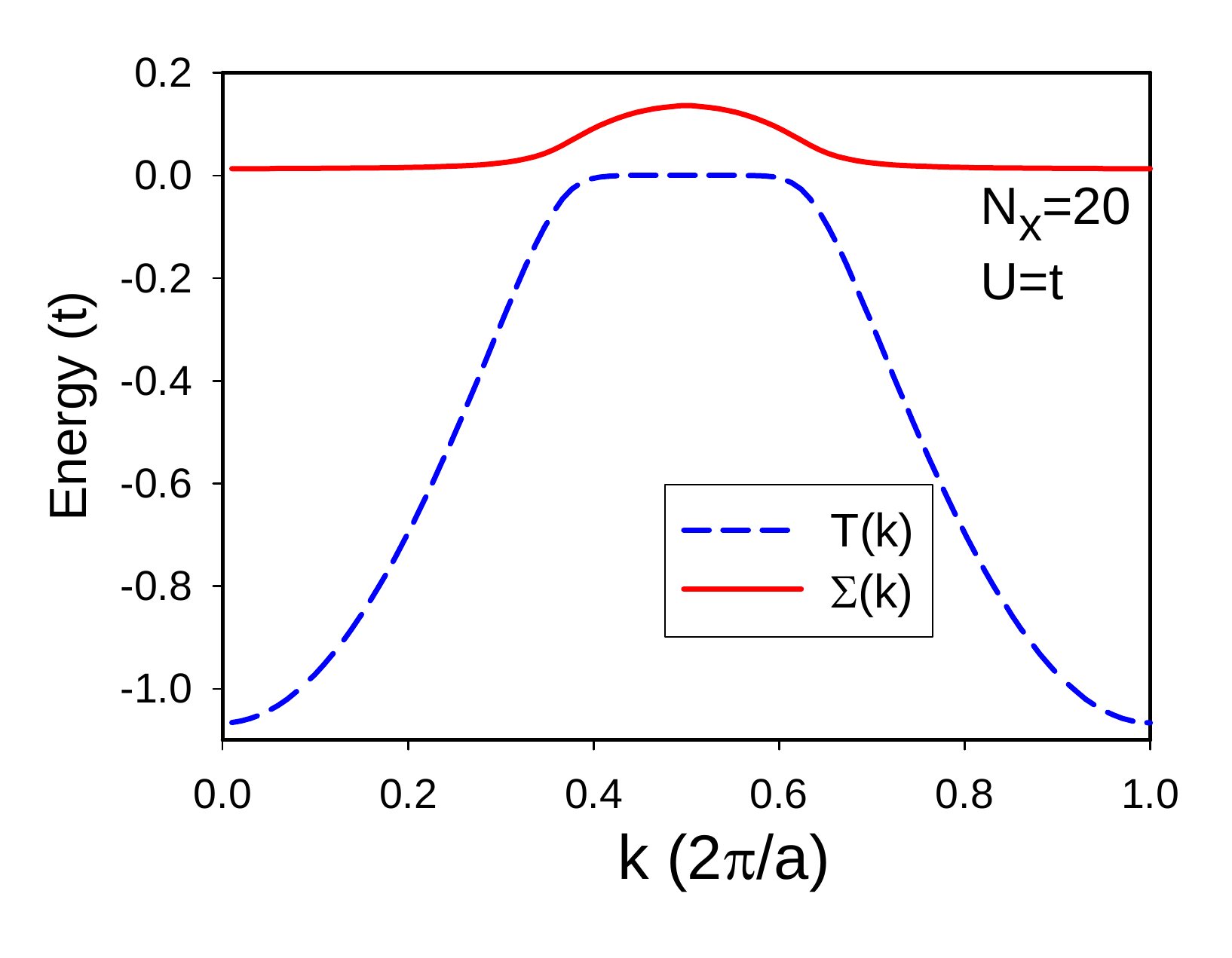}
\caption{Selfenergy and tunneling amplitude between states centered in the left and right  edges of a nanoribbon as function of the wavevector $k$. In the calculation we use  $N_x$=20 and $U=t$. } 
\label{Figure2}
\end{figure}

\par
\noindent
{{\it 2$\times$2 Effective Hamiltonian.-} 
Although the charge density is uniform along and across the nanoribbon, the spin polarization produces spin-dependent electric polarizations\cite{Fernandez-Rossier:2008aa}. 
To a great degree of precision the magnetic properties of zigzag graphene nanoribbons can be described by restricting the Hilbert space to the  highest energy valence band, $|k,->$ and the lowest energy conduction band, $|k,+>$ of the non interacting, $U=0$,
Hamiltonian\cite{Fernandez-Rossier:2008aa,Jung:2009aa}.   The wave functions of these states are even and odd combinations of the $\pi$ orbitals across the nanoribbon. As the electric and magnetic properties of the nanoribbon are associated with localization of charge at the edges, it is appropriated  to use
a local base of the form  
\begin{equation}
|k,L (R)>= \frac 1 {\sqrt{2}} \left( |k,+> \pm |k,->  \right )
\end{equation}
in this basis the self-consistent Hamiltonian for each spin orientation takes the form,
\begin{equation}
H_{\sigma} (k) =\left ( \begin{array}{cc} -\sigma  \Sigma (k) & T(k) \\ T(k) & \sigma  \Sigma (k) \end{array} \right )
\label{H22} 
\end{equation}
where $\sigma =\pm1$ for for spins pointing up or down respectively, $T(k)$ is the $k$-dependent hopping between left and right located states and
$\Sigma (k)$ is the exchange self energy  that has opposite sign for states located on opposite  edges or with opposite spin. Both the hopping amplitude and the self energy are real quantities which are obtained by solving self-consistently the Hubbard Hamiltonian.
In Fig.\ref{Figure2} we plot $T(k)$ and $\Sigma (k)$ for a particular zigzag graphene nanoribbon.
From the eigenvectors and eigenvalues of the previous Hamiltonian, the  spin-dependent electric dipoles in the transverse direction, $\hat x$, of the nanoribbon, take the form,
\begin{equation}
{\cal P}_{\sigma} = e \sigma \sum _{k} \Sigma (k) <k,L|x|k,R> \, .
\end{equation}
As commented above the system is not ferrolelectric and the sum of the spin-dependent electric dipoles vanishes, 
${\cal P}_{\uparrow}+{\cal P}_{\downarrow}=0$.

The form of the two Hamiltonians, Eq.\ref{H22}, indicates the degeneracy of the ground state; by reversing the spin orientation of the full system,  another degenerate ground state appears, where the spin polarization at the edges and the spin-dependent electric polarization  are reversed, see Fig.\ref{Figure1}.  We characterized the two degenerate ground states by ${\cal T }$=${\rm sign} {\cal P} _{\uparrow}$.  
The index ${\cal T}$ shows the spin polarization  of the dipole generated in the ribbon by a transversal electric field\cite{Fernandez-Rossier:2008aa}.
The degeneracy of the ground state reflects the broken symmetry of the ground state in the spin sector. 
Associated with this degeneracy  there  exist topological excitations, in this case solitons.  
For each spin orientation, the two degenerate ground states  have the bands inverted  
and therefore there will be  two  topological protected states, one for each spin orientation,  at the interface between  domains with opposite ${\cal T}$.  

Both Hamiltonians, Eq.\ref{H22},  satisfy the anticommutation relation $\tau_y H_{\sigma} (k) \tau _y$=-$H_{\sigma} (k)$, being $\tau _y$  a Pauli matrix. A consequence of this symmetry is that the spectrum of $H_{\sigma}$ is electron hole symmetric, any eigenstate $|\psi> $with energy $\epsilon$ has a conjugate state $\tau _y |\psi>$ with energy $-\epsilon$. Because the electron hole symmetry,  the topological protected states  should be placed at the middle of the gap and get   zero energy.  
Half of the spectral weight of the mid gap state comes from the conduction band and the other from the valence band, therefore
when the chemical potential is above (below)  zero energy,  the soliton  carries a charge -$e/2$ ($e/2$)\cite{Jackiw:1976aa,Heeger:1988aa}. Considering the two spin orientations, the topological excitation in ZZGN's consists of two $e/2$ charged solitons, and carry a total charge $e$.
The connection between topological defects and electric charge suggests that solitons can be the relevant charge excitation in 
zigzag graphene nanoribbons. Then whenever   adding (subtracting )  charge to the system an array of solitons  can be formed, creating a solitonic phase.  The distance between solitons is  the inverse of the density of extra charge per unit length in the ribbon.

\begin{figure}[htbp]
\includegraphics[width=8.5cm,clip]{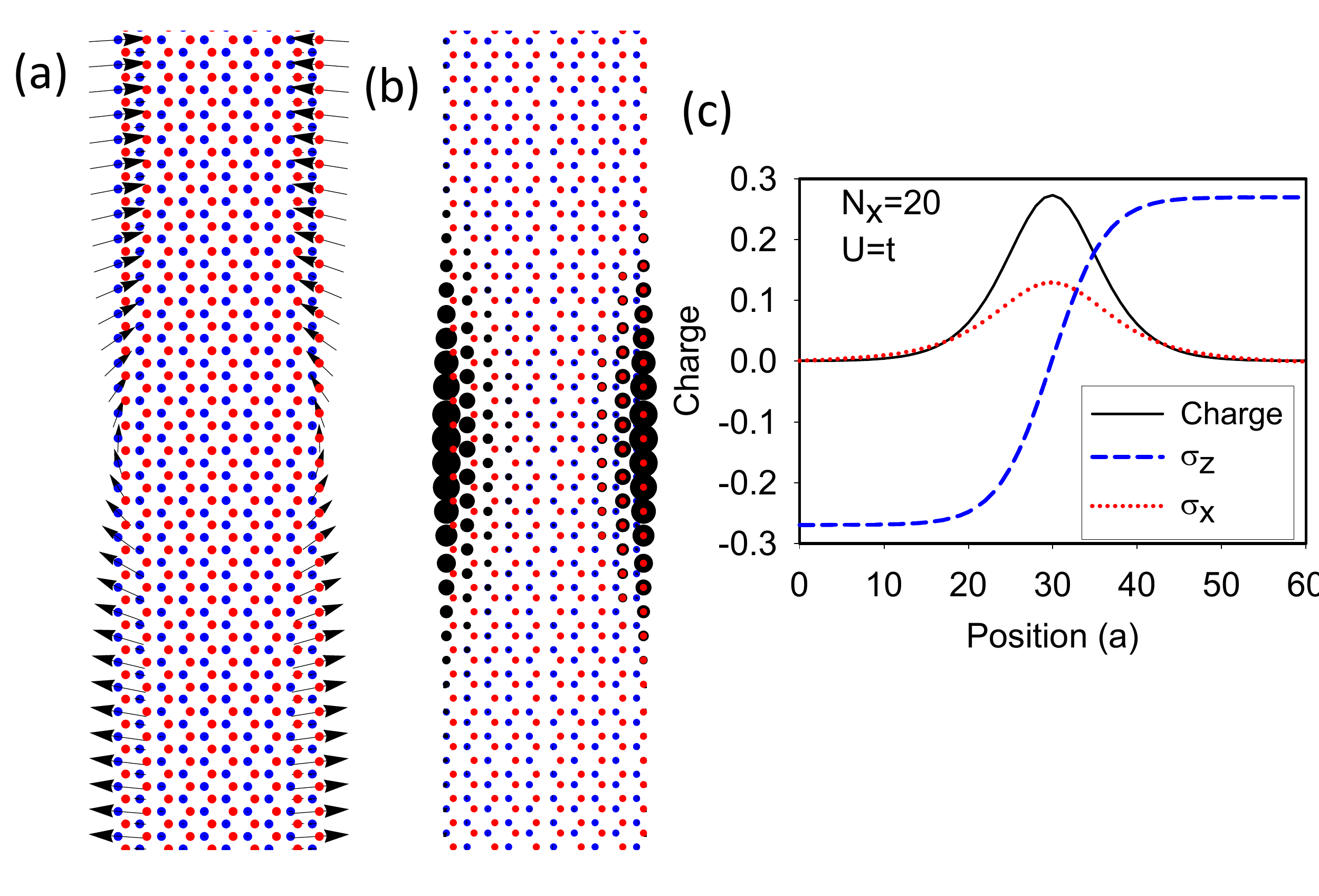}
\caption{(a) Local spin polarization and (b) excess of charge near a domain wall separating two degenerate 
gapped ground states.  In (c) we plot the same quantities on the outermost atom in the left edge of the ribbon as function
of the position along the ribbon. In the right edge  $\sigma _z$ changes sign, whereas $\sigma _x$ does not. In (a) the orientation of the spins are rotated to make the figure clearer. In the calculations we use $N_x$=20, $N_y$=120 and $U=t$. } 
\label{Figure3}
\end{figure}
\par
\noindent
{\it Numerical results.}- To verify  and quantify this proposal, we compute the energy and the electric and magnetic properties
 of  a ZZGN in presence of an extra number of electrons.
Because of the electron-hole symmetry existing in the system, the  calculations are restricted just to the case of doping with electrons.
We consider a periodic structure along the nanoribbon, with a supercell containing $N_y$ repetitions of the minimum  unit cell, so that the unit cell contains $N_y \times N_x$ carbon atoms. 
Because of the use of periodic boundary conditions the solitons in the unit cell always appears by pairs.  
We add a number $n_{extra}$ of electrons to the unit cell and because of the one dimensional nature of the system
 the excess of charge   is  expressed  as density of electrons  per unit length in the ribbon, 
$\delta n$=$n_{extra}/(N_y a)$. 
By solving self-consistently the 
Hubbard  Hamiltonian in the unrestricted Hartree-Fock approximation, we obtain the energy and the 
spin and charge spatial distribution in the ribbons as function of the electron density. 
The solutions converge to the solitonic phase when imposing the initial guess with the appropriated spin-spatial  distribution.
In Fig.\ref{Figure3}  we show the spatial spin polarization (a) and  charge distribution (b) for a soliton separating two domains with
opposite ${\cal T}$. Crossing the domain wall, the spin polarization in the left edge rotates from pointing in the $+\hat z$ direction to pointing in the $-\hat z$ direction, 
acquiring the electron spin polarization a 
small $\hat x$-component. In the right edge, the spin polarization 
in the $\hat z$ direction has opposite sign, whereas the $\hat x$-component of the spin polarization, in both edges, are parallel. 

\begin{figure}[htbp]
\includegraphics[width=8.5cm,clip]{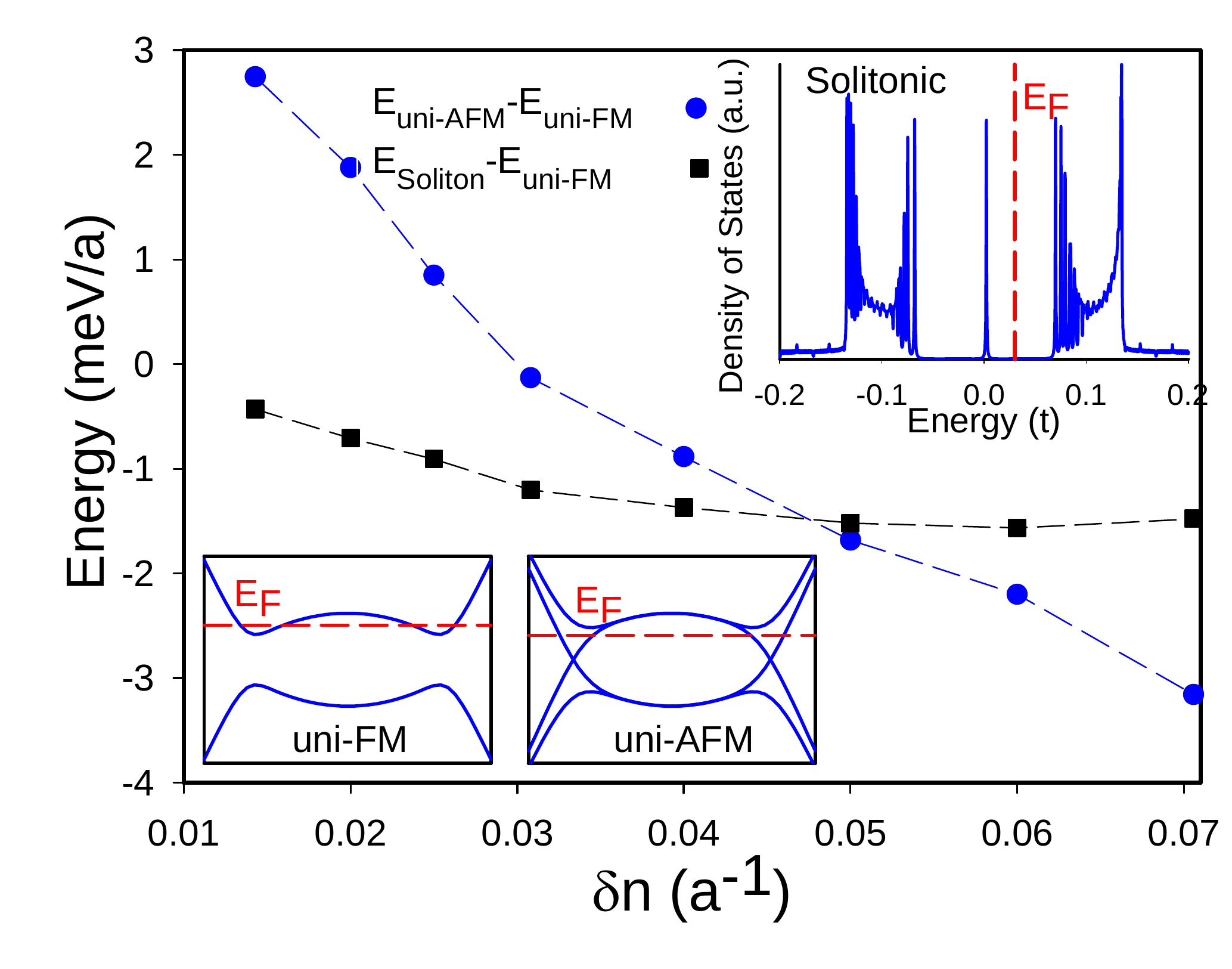}
\caption{Energy per unit length of different phases of doped zigzag graphene nanoribbons. The energies are referred to that of the uni-FM phase. At low doping, $\delta n <0.05/a$ the solitonic phase is the ground state of the system. In the lower panels,  we show schematically  the band structures of the uniform phases and the position of the Fermi energy. In the inset in the upper part of the figure, we plot the density of states of the solitonic phase. The zero energy peak corresponds to the solitons. The calculations are done with $N_x$=20, $U=t$ and $t$=2.7eV. } 
\label{Figure4}
\end{figure}

In order to verify  that the solitonic phase is the ground state at low densities we compare its  energy with the energies of phases with uniform distribution of charge and spin polarization along the nanoribbon. In particular we compare with uniform phases with ferromagnetic (uni-FM) or antiferromagnetic (uni-AFM) coupling between the edges\cite{Jung:2009ab}.  Strictly, the spin polarization  of the uniform
doped phases
is not collinear and the edge polarizations are  slightly canted with respect the FM and AFM order. 
In Fig.\ref{Figure4}  we plot the the total energy difference per unit length between the solitonic and the uni-FM and uni-AFM phases, for a nanoribbon with $N_x$=20. The energy is referred to the energy of the uni-FM phase. From the results shown in this figure, we conclude that the solitonic phase is the ground state of the nanoribbon at densities lower than $\delta n \sim$0.05$/a$. At these densities the system prefers to create domains with opposite ${\cal T}$ and accommodate the extra carriers at the solitons that appear at the domain walls. 

\par
\noindent
{\it Relation with previous works and discussion.}
Using density functional theory\cite{Sawada:2009aa} or tigh-binding Hamiltonians\cite{Dutta:2012aa,Li:2016aa,Jung:2009ab}, previous theoretical works have 
found that   zigzag gaphene nanoribbons  become ferromagnetic when doped. Using  the same Hubbard model than us, Jung and MacDonald\cite{Jung:2009ab} find the  transition from the uni-FM to the uni-AFM at a density $\delta n \sim$ 0.03$/a$. However, all these calculations do not allow the modulation of the charge and spin along the ribbon, and therefore did not find any clue for  the existence of the solitonic phase.
The presence of a solitonic phase in zigzag graphene nanoribbons should be detected in transport experiments, as a significant enhancement in the electrical transport in the middle of the energy gap. Also in the case of solitons trapped by defects or impurities, individual solitons could be visualized 
by scanning tunneling microscopy experiments. 
The typical  size of the solitons is  of some graphene lattice parameters, $\sim 20a$, and  their existence
would require nanoribbons with magnetic correlation length larger than this size.  Recent calculations by Yazyev {\it et al.}\cite{Yazyev:2008aa}
have shown that,  at low temperatures, the correlation length in ZZGN's could be as larger as 300$\AA$ at 10$K$, and this means that solitons
can be observed at low temperatures. 
The calculations presented in this work are done for ribbons with $N_x$=20, we have checked that solitonic phases appear 
in wider ribbons. The conditions for the appearance of solitons is the existence of an energy gap,
and in  ZZGN's the gap scales as $W^{-1}$\cite{Jung:2009aa}. 
The value of the gap increases with the value of the Hubbard interaction $U$, in our calculations we use a value of $U$=$t$, that reproduces the 
gap obtained in density functional calculations. Non local Coulomb interactions can modify the value of $U$, and recently truly {\it ab initio} 
calculations\cite{Wehling:2011aa,Schuler:2013aa} have reported  a value of $U\sim 2 t$, that makes the magnetic ground state of the ZZGN more stable and, therefore more robust the existence of a solitonic phase .

In summary, we have shown that the low energy charge excitations in doped graphene zigzag nanoribbons are solitons.
These solitons appear because of the degeneracy of the magnetic broken symmetry ground state of the ribbon. The two 
degenerate states have the electronic bands inverted, and when joining two opposite  magnetic domains a topological soliton will
appear at the interface. The solitons are the low energy charge excitations in the system, and when doping the extra charge will create
magnetic domains and will accommodate at the solitons.

\setlength{\parskip}{0.1cm}
\par
\noindent
{\it Acknowledments.}-
M.P.L.-S. acknowledges financial support by the Spanish
MINECO Grant No. FIS2014-57432-P, the European Union
structural funds and the Comunidad Aut\'onoma de Madrid
MAD2D-CM Program (S2013/MIT-3007). LB  acknowledges financial support by the Spanish
MINECO Grant No. FIS2015--64654-P.
\setlength{\parskip}{0.cm}
 \end{document}